\documentclass{osa-article}

\journal{oe}
\articletype{Research Article}

\usepackage{booktabs}
\usepackage{bbm}
\usepackage{amsmath}
\DeclareMathOperator*{\argmax}{\mathrm{argmax}}

\begin{document}

\title{End-to-end optimized transmission over dispersive intensity-modulated channels using bidirectional recurrent neural networks}

\author{Boris~Karanov,\authormark{1,2,*} Domani\c{c}~Lavery,\authormark{1} Polina~Bayvel,\authormark{1} and Laurent~Schmalen\authormark{2,3}}

\address{\authormark{1}Optical Networks Group, Department of Electronic and Electrical Engineering, University College London (UCL), WC1E 7JE London, UK\\
\authormark{2}Nokia Bell Labs, 70435 Stuttgart, Germany\\
\authormark{3}\emph{now with} Karlsruhe Institute of Technology (KIT), Communications Engineering Lab, Karlsruhe, Germany}

\email{\authormark{*}boris.karanov.16@ucl.ac.uk}

\begin{abstract}
We propose an autoencoding sequence-based transceiver for communication over dispersive channels with intensity modulation and direct detection (IM/DD), designed as a bidirectional deep recurrent neural network (BRNN). The receiver uses a sliding window technique to allow for efficient data stream estimation. We find that this sliding window BRNN (SBRNN), based on end-to-end deep learning of the communication system, achieves a significant bit-error-rate reduction at all examined distances in comparison to previous block-based autoencoders implemented as feed-forward neural networks (FFNNs), leading to an increase of the transmission distance. We also compare the end-to-end SBRNN with a state-of-the-art IM/DD solution based on two level pulse amplitude modulation with an FFNN receiver, simultaneously processing multiple received symbols and approximating nonlinear Volterra equalization. Our results show that the SBRNN outperforms such systems at both 42 and 84\,Gb/s, while training fewer parameters. Our novel SBRNN design aims at tailoring the end-to-end deep learning-based systems for communication over nonlinear channels with memory, such as the optical IM/DD fiber channel.
\end{abstract}

\section{Introduction}
Deep learning techniques, enabling the approximation of any nonlinear function~\cite{Goodfellow}, allow us to design communication systems by realizing the optimization of the transceiver in a single end-to-end process including the complete chain of transmitter, communication channel and receiver. Such systems, implemented as a single deep neural network, have the prospect of achieving an optimal end-to-end performance and have attracted attention in communication scenarios where the optimum pair of transmitter and receiver or optimum processing modules are not known or not available because of complexity reasons. Recently, this approach has been introduced and experimentally verified for both wireless~\cite{O'Shea,Doerner} and optical fiber communications~\cite{Karanov,Karanov_2,Li,Jones}. Applied to intensity modulation/direct detection (IM/DD) fiber-optic systems, this novel design outperformed both in simulation and experimentally ubiquitously deployed pulse amplitude modulation (PAM2/PAM4) schemes with specific, conventional linear equalizers~\cite{Karanov, Karanov_2}. The IM/DD optical communication channel is nonlinear and has memory due to the effects of intensity detection by a photodiode~(PD) and fiber chromatic dispersion (CD)~\cite{Agrawal}. 

In~\cite{Karanov,Karanov_2}, we implemented a block-based end-to-end deep learning optical fiber system, each block~(symbol) representing an independent message of a few data bits only. This allowed us to design the system as a feed-forward neural network (FFNN) and enabled efficient transceiver implementation by parallel processing of the blocks. However, the end-to-end FFNN system is inherently unable to compensate for CD outside of the block, as connections between neighboring blocks are not included in the design structure. Thus, the inter-block interference is treated as extra noise and, as a consequence, the achievable performance, in terms of CD that can be compensated and hence transmission distance, of such systems is limited by the block size.

In this work, we address the limitations of the FFNN design for communication over nonlinear channels with memory by implementing \emph{sequence-based} end-to-end deep learning transceivers using recurrent neural networks (RNN)~\cite{Goodfellow,Rumelhart}. RNNs have been recently demonstrated as a viable receiver-only signal processing solution in passive optical networks (PON)~\cite{Ye}. We use RNNs to design an end-to-end optimized fiber-optic system resilient to the nonlinearities and inter-symbol interference (ISI) present in IM/DD communications over dispersive channels. More specifically, since CD causes ISI from both preceding and succeeding symbols, we employ in our design bidirectional RNNs (BRNN)~\cite{Schuster}. We operate the trained BRNN transceivers in a sliding window sequence estimation scheme (SBRNN), which allows us to estimate the data stream efficiently~\cite{Farsad,Farsad_2}. In contrast to~\cite{Farsad,Farsad_2}, where the neural network processing is only at the receiver side, in our work we employ BRNN structures at both transmitter and receiver to allow end-to-end optimization of the transmission over the communication channel. Two variations of the recurrent cell in the RNN structure are examined, a straightforward \emph{vanilla} concatenation as well as the long short-term memory gated recurrent unit structure (LSTM-GRU), specifically designed to handle long term dependencies in the sequence~\cite{Goodfellow,Hochreiter,Cho,Ravanelli}. We find that the LSTM-GRU design has a slightly superior bit error rate (BER) performance compared to the vanilla SBRNN, however at a higher computational cost. 

Our study shows that both SBRNN systems, specifically designed to handle the channel memory, can significantly outperform the previous end-to-end FFNN. Moreover, we compare the SBRNNs with PAM2 systems deploying a large FFNN at the receiver, which processes multiple received symbols also in a sliding window scheme. Such an FFNN receiver has been shown to perform on par with Volterra equalization and it has been considered in low-cost IM/DD links such as PONs~\cite{Bell_labs_OFC2019_Volterra, Lyubomirsky}. Our results show that SBRNN outperforms the reference system, achieving information rates of 42\,Gb/s below the 6.7\% hard-decision forward error correction (HD-FEC) threshold at distances of 70\,km. We also show that with a sufficiently large processing window, the SBRNN can achieve 84\,Gb/s at 30\,km.

\section{End-to-end SBRNN system design}

\begin{figure}[t!]
\centering
\includegraphics[width=\textwidth,keepaspectratio]{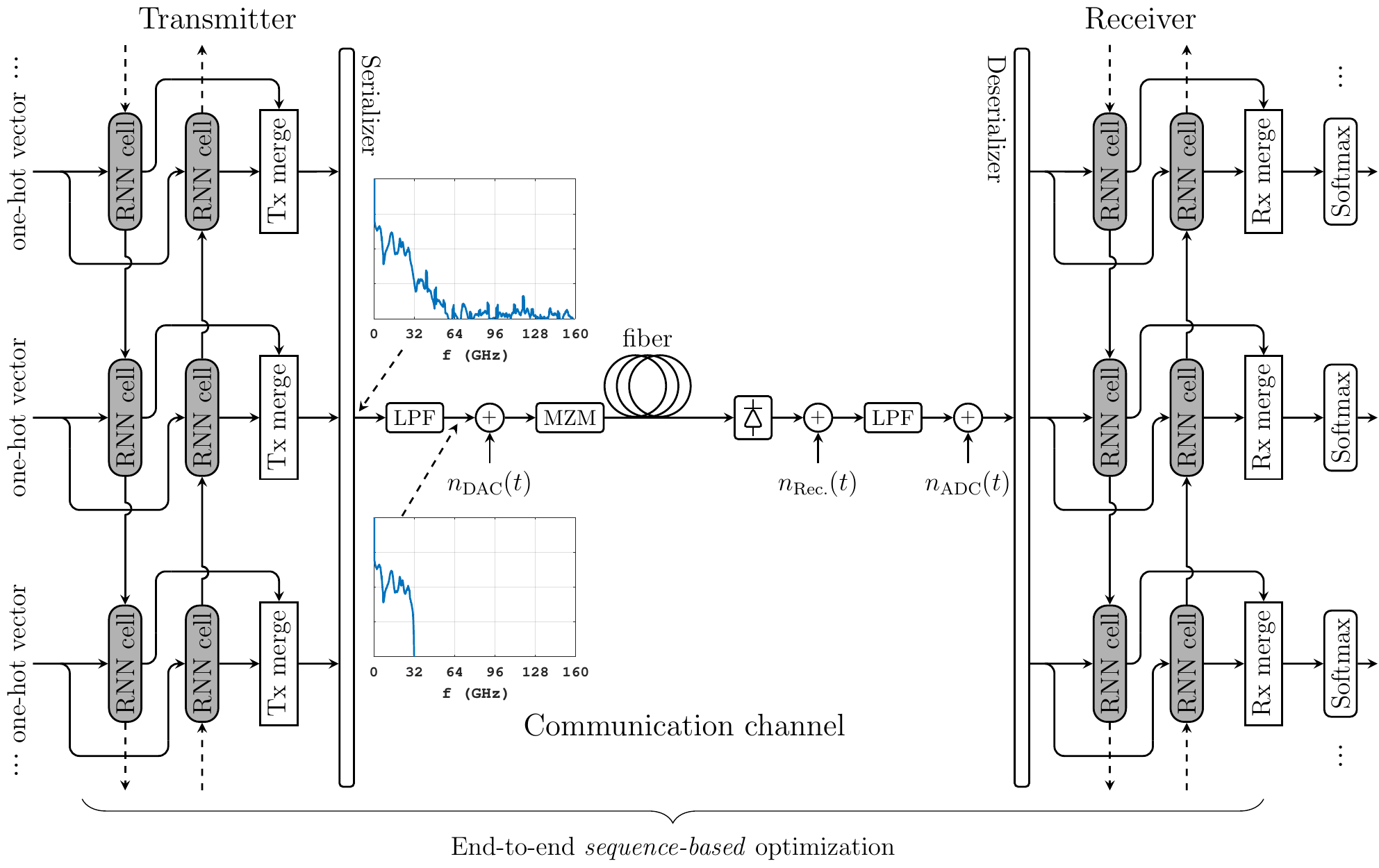}
\caption{\label{fig:Schematics} Schematic of the IM/DD optical fiber communication system implemented as a bidirectional deep recurrent neural network. Optimization is performed between the stream of input messages and the outputs of the receiver, thus enabling end-to-end optimization via deep learning of the complete system. Inset figures show the transmitted signal spectrum both at the output of the neural network and before the DAC.}
\end{figure}

The complete optical communication system is implemented as an end-to-end deep neural network as in~\cite{Karanov}, following the idea of~\cite{O'Shea}. In this work, we use a bidirectional deep recurrent neural network and Fig.~\ref{fig:Schematics} shows the full end-to-end BRNN chain of transmitter, receiver and communication channel. This section describes in detail all components of the complete design as well as the sliding window estimation scheme in which the trained transceivers are operated.
\subsection{Communication channel} \label{sec:Communication_channel}
Similar to~\cite{Karanov,Karanov_2}, we consider an optically un-amplified IM/DD link, a preferred solution in many low-cost short-reach applications. The fiber dispersion and the nonlinearity stemming from the square-law PD opto-electrical conversion are the dominant limiting factors in such a communication channel. As a result of the joint effects of dispersion and square-law detection, the IM/DD channel is nonlinear and has memory, meaning that inter-symbol interference is induced from both preceding and succeeding symbols. Therefore, sequence processing is required for communication over such channels. In this work, we assume a channel model that includes low-pass filtering (LPF) at transmitter and receiver to reflect current hardware limitations, digital-to-analog and analog-to-digital converters (DAC/ADC), Mach-Zehnder modulation (MZM), PD, electrical amplification noise and optical fiber transmission. For details we refer the interested reader to~\cite{Karanov}. The optical fiber is modeled as an \emph{attenuating} and dispersive medium, neglecting nonlinearities due to the Kerr effect. The MZM is modeled by its sinusoidal electrical field transfer function. The quantization noise from the DAC/ADC ($n_{\text{DAC}}(t)$ and $n_{\text{ADC}}(t)$) is modeled as additive and uniformly distributed, with variance determined by the effective number of bits (ENOB). Assuming the PAM2 system examined in Sec.~\ref{sec:Performance} and detailed in~\cite{Karanov} as a reference, we use the experimentally obtained received SNR of 19.41\,dB at 20\,km~\cite{Karanov,Karanov_2} to estimate the variance $\sigma_r^2$ of the additive receiver white Gaussian noise $n_{\text{Rec.}}(t)$. In contrast to~\cite{Karanov,Karanov_2}, our channel model includes fiber attenuation and thus $\sigma_r^2$ is constant as a function of the transmission distance. We apply the estimated value of $\sigma_r^2=2.455\cdot 10^{-4}$ for all examined systems. As the fiber model includes attenuation (0.2\,dB/km), this will yield different effective SNRs at each transmission distance.

\subsection{Transceiver design} \label{transceiver_design}
Optical fiber dispersion introduces ISI effects from both preceding and succeeding symbols, which we take into account by considering an end-to-end design based on bidirectional recurrent structures. The BRNN at the transmitter encodes a message $m\in\{1,\ldots, M\}$ independently drawn from a set of $M$ total messages into a vector of $n$ transmit samples. The message $m$ is represented as a one-hot vector $\mathbf{1}_{m}\in{\mathbb{R}}^{M}$ (which contains a ``1'' at position $m$ and zeros everywhere else) and fed for processing as input $\mathbf{x}_{t}$ to the transmitter RNN cells in both directions. After propagation through the communication channel, the received samples are applied for processing in the receiver section of the BRNN.

\begin{figure}[t!]
\centering
\includegraphics[width=0.35\textwidth, keepaspectratio=true]{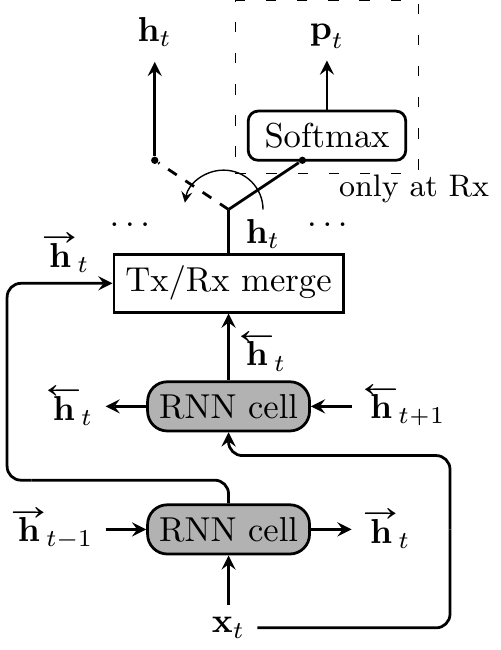}
\caption{\label{fig:BRNN_Schematics} Bidirectional RNN schematic. The final transmitter/receiver outputs are obtained by merging the outputs of the forward and backward passes. The transmitter outputs are sent through the communication channel, while \emph{softmax} is applied to the receiver outputs resulting in probability vectors $\mathbf{p}_{t}$, utilized in the sliding window estimation.}
\end{figure}

Figure~\ref{fig:BRNN_Schematics} shows a schematic of the BRNN concept, which we use as building blocks of the proposed transceiver. In the forward direction, an input $\textnormal{\textbf{x}}_{t}$ at time $t$ is processed by the RNN cell together with the previous output $\textnormal{\textbf{h}}_{t-1}$ to produce an updated output $\textnormal{\textbf{h}}_{t}$. This procedure is performed across the full data sequence. To account for the interference from the succeeding symbols, the structure is repeated in the backward direction as well. At the transmitter, we input \emph{one-hot vectors} and average the outputs of the RNN cells at the same time instance in both directions in the \emph{Tx merge} block. The transmitter output at a time instance $t$ is
$\mathbf{h}_{t} = \frac{1}{2}\left({\overrightarrow{\mathbf{h}}}_{t}+{\overleftarrow{\mathbf{h}}}_{t}\right)$,
 where ${\overrightarrow{\mathbf{h}}}_{t}$ and ${\overleftarrow{\mathbf{h}}}_{t}$ are the outputs of the RNN cells in the forward and the backward directions, respectively. 
 
 At the receiver, we concatenate the two RNN cell outputs in the \emph{Rx merge} block and at time $t$, the output of the BRNN is $\mathbf{h}_{t} = \begin{pmatrix} \overrightarrow{\mathbf{h}}_t^T & \overleftarrow{\mathbf{h}}_{t}^T\end{pmatrix}^T$,
where ${}^{T}$ denotes the matrix transpose operation. A \emph{softmax} layer is applied to the output of the receiver BRNN to obtain probability vectors $\mathbf{p}_t = \mathop{\textrm{softmax}}(\mathbf{W}_{\text{softmax}}\mathbf{h}_t^T + \mathbf{b}_{\text{softmax}})$ with $\mathbf{p}_t\in{\mathbb{R}}^{M}$, $\mathbf{W}_{\text{softmax}}\in{\mathbb{R}}^{M\times 4M}$ and $\mathbf{b}_{\text{softmax}}\in{\mathbb{R}}^{M}$. The \emph{softmax} activation function is defined as $\mathbf{y}= \mathop{\textrm{softmax}}(\mathbf{x})$ with
\begin{equation}
y_{i}=\frac{\exp(x_i)}{\sum\limits_{j}\exp(x_j)}.
\end{equation}

\begin{figure}[t!]
\centering

\includegraphics[width=0.32\textwidth, keepaspectratio=true]{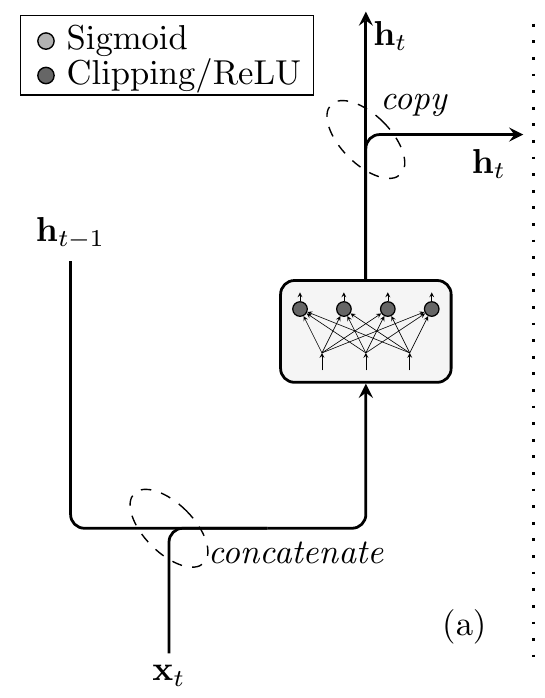}
\includegraphics[width=0.4\textwidth, keepaspectratio=true]{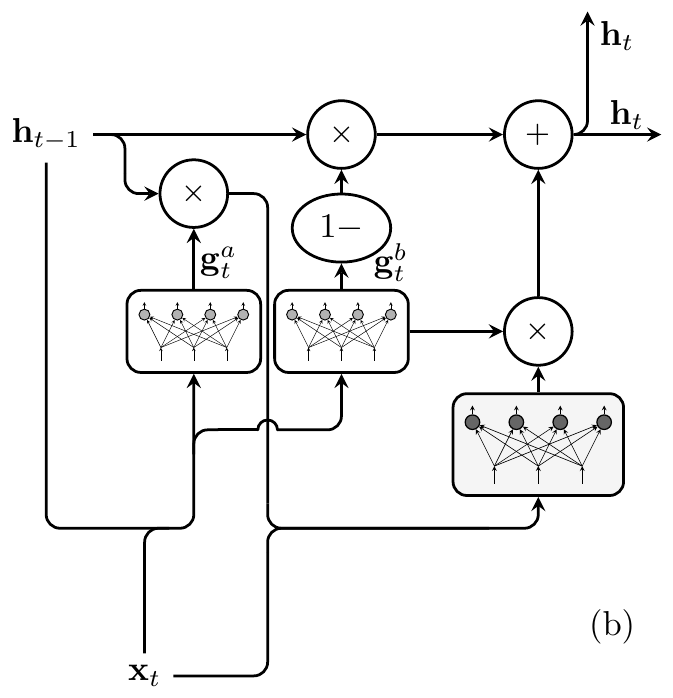}
\caption{\label{fig:Cell_Schematics} Schematic of: (a) a vanilla RNN cell, (b) an LSTM-GRU cell. Lines merge when their content is concatenated and diverge when it is copied.}
\end{figure}

Our work examines two RNN cells, a \emph{vanilla} cell, consisting of a simple concatenation, as well as a variant of the long-short term memory (LSTM) gated recurrent unit (GRU)~\cite{Goodfellow,Cho,Ravanelli}. We first describe the \emph{vanilla} RNN cell, shown in Fig.~\ref{fig:Cell_Schematics}(a). The current output of the cell can be expressed as
\begin{equation}
\mathbf{h}_{t}=\alpha\left(\mathbf{W}\begin{pmatrix} \mathbf{x}_t^T & \mathbf{h}_{t-1}^T\end{pmatrix}^T +\mathbf{b}\right),
\end{equation}
where $\alpha$ is the corresponding activation function at the transmitter/receiver. Note that in contrast to the feed-forward neural networks~\cite[Sec. II, Eq. (1)]{Karanov}, the recurrent neural networks process the concatenation of the current input with the previous output. This manifests the application of RNNs in communication systems, such as the optical IM/DD, where sequence processing is required.
At the transmitter, the activation function limits the outputs to $[0;\pi/4]$ to allow for linear MZM operation. For this reason, we use the clipping activation function~\cite{Karanov} at the transmitter expressed as
\begin{equation}
\label{eq:ModifiedActivation}
\alpha_{\text{Tx}}(\mathbf{x})=\alpha_{\tiny{\text{ReLU}}}\left(\mathbf{x}\right)-\alpha_{\tiny{\text{ReLU}}}\left(\mathbf{x}-\frac{\pi}{4}\right),
\end{equation}
where $\mathbf{y}=\alpha_{\tiny{\text{ReLU}}}(\mathbf{x})$ is the ReLU function applied element-wise, i.e., $y_{i}=\max(0,x_i)$  ~\cite{Goodfellow,Nair}. At the receiver, we employ the ReLU activation, i.e., $\alpha_{\text{Rx}} = \alpha_{\tiny{\text{ReLU}}}(\mathbf{x})$.
The nonlinear processing network inside the cell consists of a single neural network (NN) layer with parameters  $\mathbf{W}_{}\in{\mathbb{R}}^{{n \times (M+n)}}$ and $\mathbf{b}_{}\in{\mathbb{R}}^{n}$ (transmitter) or $\mathbf{W}_{}\in{\mathbb{R}}^{{2M \times (n+2M)}}$ and $\mathbf{b}_{}\in{\mathbb{R}}^{2M}$ (receiver), respectively.

The alternative \emph{LSTM-GRU} cell is shown in Fig.~\ref{fig:Cell_Schematics}(b). Compared to the \emph{vanilla} cell, it consists of two additional single layer memory gates $\mathbf{g}_{t}^{a}$ and $\mathbf{g}_{t}^{b}$. Their function is to decide which elements of the current input/previous output should be preserved for further processing. The current output of the LSTM-GRU cell is obtained as
\begin{equation}
\mathbf{g}_{t}^{a}=\sigma\left(\mathbf{W}_{1}\begin{pmatrix} \mathbf{x}_t^T & \mathbf{h}_{t-1}^T\end{pmatrix}^T+\mathbf{b}_{1}\right), 
\end{equation}
\begin{equation}
\quad \mathbf{g}_{t}^{b}=\sigma\left(\mathbf{W}_{2}\begin{pmatrix} \mathbf{x}_t^T & \mathbf{h}_{t-1}^T\end{pmatrix}^T+\mathbf{b}_{2}\right),
\end{equation}
\begin{equation}
\mathbf{h}_{t}=(1-\mathbf{g}_{t}^{b})\odot\mathbf{h}_{t-1}+\mathbf{g}_{t}^{b}\odot\alpha\left(\mathbf{W}_{3}\begin{pmatrix} \mathbf{x}_t^T & \left(\mathbf{g}_{t}^{a}\odot\mathbf{h}_{t-1}\right)^T\end{pmatrix}^T+\mathbf{b}_{3}\right),
\end{equation}
where $\odot$ denotes the Hadamard product (element-wise multiplication of vectors) and $\mathbf{W}_{i}$ and $\mathbf{b}_{i}$ are the layers' weights and biases, whose sizes are identical to the size of the weight and bias of the \emph{vanilla} cell. The function $\sigma(\cdot)$ denotes the sigmoid activation function~\cite{Goodfellow}, i.e., $\mathbf{y}=\sigma(\mathbf{x})$ with 
\begin{equation}
y_{i}=\frac{1}{1+\exp(-x_i)}.
\end{equation}

We use the cross entropy between the one-hot vector inputs and the output probability vectors as a loss function and perform a stochastic gradient descent optimization of the neural network parameters~\cite{Goodfellow}\cite[Sec. II]{Karanov}, applying the Adam algorithm~\cite{Kingma}. All numerical results are generated with the deep learning library TensorFlow~\cite{Tensorflow}.
The proposed systems are trained in the following way: A set of $B\!=\!250$ different sequences of $T_{\textnormal{train}}=\!10^{6}$ random input messages $m_{i,j}$ is generated, with $i\in\{1,\ldots,B\}$ and $j\in\{1,\ldots,T_{\textnormal{train}}\}$, and to begin, the outputs ${\overrightarrow{\mathbf{h}}}_{t-1}$ and ${\overleftarrow{\mathbf{h}}}_{t+1}$ in the forward and backward passes of the BRNN are initialized to $\mathbf{0}$. A total of 100 000 optimization iterations is performed. At optimization step $s$, the batch of messages $m_{i,(s-1)W+1},\ldots,m_{i,sW}$ is processed by the transmitter BRNN to obtain the blocks $\mathbf{h}_{i,(s-1)W+1},\ldots,\mathbf{h}_{i,sW}$. Prior to feeding them to the channel, the blocks are transformed into long series  $\mathbf{h}_{1,(s-1)W+1},\ldots,\mathbf{h}_{1,sW},\mathbf{h}_{2,(s-1)W+1},\ldots,\mathbf{h}_{2,sW},\ldots,\mathbf{h}_{B,(s-1)W+1}\ldots,\mathbf{h}_{B,sW}$. At the receiver, this transformation is reversed and the received blocks $\mathbf{y}_{i,(s-1)W+1},\ldots,\mathbf{y}_{i,sW}$ are applied to the BRNN to obtain output probability vectors $\mathbf{p}_{i,(s-1)W+1},\ldots,\mathbf{p}_{i,sW}$. The cross entropy loss between inputs and outputs is averaged over the whole batch and a single iteration of the optimization algorithm is performed.  Note that every 100 steps of the optimization, we re-initialized the outputs ${\overrightarrow{\mathbf{h}}}_{t-1}$ and ${\overleftarrow{\mathbf{h}}}_{t+1}$ in the forward and backward passes of the BRNN to $\mathbf{0}$ to avoid getting trapped in a local minimum. 
 In Sec.~\ref{sec:Performance} we investigate the impact of the processing window size {$W$} on the system performance. It {should be mentioned} that convergence of the loss in the trained models was obtained within the 100 000 iterations, used as a stopping criterion, and thus in our simulation we ran a single epoch over the training data.
\subsection{Sliding window sequence estimation technique}
\begin{figure}[t!]
\centering

\includegraphics[]{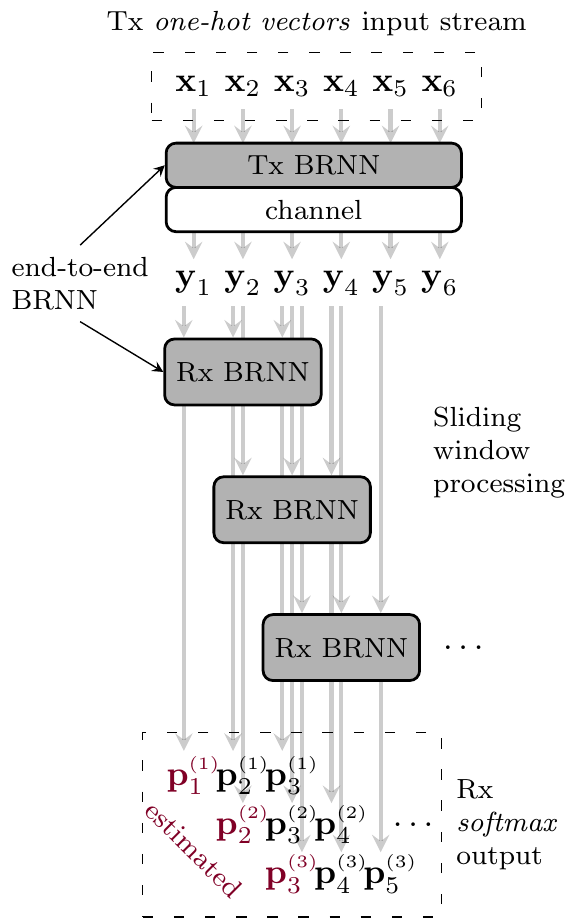}

\caption{\label{fig:SBRNN_Schematics} Schematic of the sliding window sequence estimation technique in which the BRNN transceiver is operated. Note that $W=3$ is chosen for illustration purposes.}
\end{figure}
The trained transceivers are employed in the sliding window estimation scheme introduced for receiver processing in~\cite{Farsad} and test results, presented throughout the manuscript, are generated using an independent set of different 2500 sequences of 1000 randomly chosen messages. Note that during the training process, we generate the random input messages using a Mersenne twister as a random number generator. To ensure the independence of training and testing data, and thus avoid learning representations of a pseudo-random sequence~\cite{Eriksson}, we use a Tausworthe generator (see~\cite{Lee} for details) in the test stage.

The schematic of operation for the proposed system using the sliding window sequence processing technique is depicted in Fig.~\ref{fig:SBRNN_Schematics}, where $W=3$ for illustration purposes. The end-to-end BRNN is represented by the blocks \emph{Tx BRNN}, \emph{channel} and \emph{Rx BRNN}, each of which was described in detail in the previous sections. For a given test sequence, the transmitter BRNN processes the full input stream of one-hot vectors $\mathbf{x}_1, \ldots, \mathbf{x}_{T+W-1}$, where $T+W-1$ is the sequence length. The obtained waveform is transmitted through the communication channel to obtain the sequence of received blocks of samples $\mathbf{y}_1, \ldots, \mathbf{y}_{T+W-1}$, which are then applied to the receiver BRNN in a sliding window algorithm described in the following. 

Given the sequence of received blocks, the Rx BRNN processes the window of $W$ blocks $\mathbf{y}_t, \ldots, \mathbf{y}_{t+W-1}$ and correspondingly outputs $W$ probability vectors $\mathbf{p}_{t}^{(t)},\ldots, \mathbf{p}_{t+W-1}^{(t)}$ for the inputs $\mathbf{x}_t, \ldots, \mathbf{x}_{t+W-1}$ before its position is shifted one time slot ahead. The final estimated output probability vector by the SBRNN for the first $W$ input one-hot vectors $\mathbf{x}_1, \ldots, \mathbf{x}_{W}$ is thus given by
\begin{equation}
\label{eq:Sliding_window_1}
\mathbf{p}_{i}=\frac{1}{i}\sum\limits_{k=1}^{i}\mathbf{p}_{i}^{(k)},\quad i=1, \ldots W,
\end{equation}
where $k$ corresponds to the iteration step of the estimation algorithm. The estimation of the probability for the inputs $\mathbf{x}_{W+1}, \ldots, \mathbf{x}_{T}$ is obtained as
\begin{equation}
\label{eq:Sliding_window_2}
\mathbf{p}_{i}=\frac{1}{W}\sum\limits_{k=i}^{i+W-1}\mathbf{p}_{i}^{(k-W+1)},\quad i=W+1, \ldots T.
\end{equation}
Note that the final $W-1$ messages $\mathbf{x}_{T+1}, \ldots, \mathbf{x}_{T+W-1}$ from the input sequence are not fully estimated by the algorithm and we do not include them in the error rate estimation. In the following we exemplify the sliding scheme using Fig.~\ref{fig:SBRNN_Schematics} as a reference. At the first estimation step ($k=1$) the receiver BRNN processes the sub-sequence of received blocks $\left(\mathbf{y}_1,\mathbf{y}_2,\mathbf{y}_3\right)$ for which the estimated probability vectors $\left(\mathbf{p}_1^{(1)},\mathbf{p}_2^{(1)},\mathbf{p}_3^{(1)}\right)$ are the output. Then the Rx BRNN slides forward to process $\left(\mathbf{y}_2,\mathbf{y}_3,\mathbf{y}_4\right)$ at the second instance and the output is $\left(\mathbf{p}_2^{(2)},\mathbf{p}_3^{(2)},\mathbf{p}_4^{(2)}\right)$. Notice that at this stage the probability $\mathbf{p}_1^{}$ is no longer under estimation and thus for input $\mathbf{x}_1$ the algorithm has obtained a final estimated output probability vector $\mathbf{p}_1^{}=\mathbf{p}_1^{(1)}$. Furthermore, step $k=2$ is the last overlap of $\mathbf{y}_2$ with the sliding processor and $\mathbf{p}_2^{}=\frac{1}{2}\left(\mathbf{p}_2^{(1)}+\mathbf{p}_2^{(2)}\right)$ is output as a final estimate for input $\mathbf{x}_2$. Similarly at the third sliding step, a final probability vector $\mathbf{p}_3^{}=\frac{1}{3}\left(\mathbf{p}_3^{(1)}+\mathbf{p}_3^{(2)}+\mathbf{p}_3^{(3)}\right)$ is estimated for the input $\mathbf{x}_3$, while at $k=4$ the decision for $\mathbf{x}_4$ will be given by $\mathbf{p}_4^{}=\frac{1}{3}\left(\mathbf{p}_4^{(2)}+\mathbf{p}_4^{(3)}+\mathbf{p}_4^{(4)}\right)$. The estimation algorithm carries on across the full sequence of received blocks, obtaining probabilities for the transmitted messages. Note that as the receiver BRNN slides through the complete received sequence, the tuples of current outputs from the forward and the backward neural network passes are also transferred across.

After a final probability vector for a given input is estimated, a decision on the transmitted message is made. A block error is counted when $m\neq \argmax(\mathbf{p})$, where $m$ is the index of the element equal to 1 in the input one-hot vector ($\mathbf{1}_{m}$). Consequently, the block error rate (BLER) for the transmitted sequence is estimated as
\begin{equation}
\label{eq:BLER}
\text{BLER}=\frac{1}{|T|}\sum\limits_{i\in{T}}{\mathbbm{1}}_{\left\{m_{i}\neq\argmax(\mathbf{p}_{i})\right\}},
\end{equation}
where {$|T|=1000-W+1$} is the number of {estimated} messages in the test sequence and $\mathbbm{1}_{\{\cdot\}}$ denotes indicator function, equal to 1 when the argument is satisfied and 0 otherwise.

Throughout the manuscript we use the bit-error rate (BER) metric to evaluate the system performance. For counting the bit errors from a detected block error, we use bit mapping that consists of assigning the Gray code to the input $m\in\{1,\ldots, M\}$. Note that this ad hoc approach is sub-optimal since the deep learning algorithm aims at minimizing the BLER and a symbol error may not result into a single bit error.
We obtain the final BER of the system as the average BER over the set of 2500 test sequences.

\def\myabst{-1ex}
\begin{table}
 \caption{\textbf{Simulations parameters}}
 \label{table1:Simulation parameters}
 \centering
 \begin{tabular}{cc}
 \toprule
 {Parameter} & {Value}\\[\myabst]
 \midrule
 {\textit{M}} & {64}   \\[\myabst]
 {\textit{n}} & {48 or 24} \\[\myabst]
 {Test sequence length} & {1000}  \\[\myabst]
 {Processing window \textit{W}} & {10} \\[\myabst]
 {DAC/ADC rate} & {84 GSa/s} \\[\myabst]
 {Simulation oversampling} & {4} \\[\myabst]
 {Simulation sampling rate} & {336 GSa/s} \\[\myabst]
 {Symbol rate} & {7 or 14 GSym/s} \\[\myabst]
 {Information rate} & {6 bits/symbol} \\[\myabst]
 {LPF bandwidth} & {32 GHz} \\[\myabst]
 {DAC/ADC ENOB} & {6}\\[\myabst]
 {Fiber dispersion parameter} & {17 ps/nm/km} \\[\myabst]
 {Fiber attenuation parameter} & {0.2 dB/km} \\[\myabst]
 {Noise variance} & {$2.455\cdot 10^{-4}$} \\[\myabst]
 \bottomrule

 \end{tabular}

\end{table}

\section{System performance validation}
\label{sec:Performance}

Table~\ref{table1:Simulation parameters} lists the simulation parameters for the investigated systems. Similar to the study of FFNN autoencoders~\cite{Karanov}, in our simulation we assume input messages from a set of $M\!=\!64$ (6 bits) which are encoded by the transmitter BRNN into a block (called a \emph{symbol}) of either $n\!=\!48$ or $n\!=\!24$ samples for the two examined systems transmitting at 42\,Gb/s and 84\,Gb/s, respectively. To account for possible signal spectral broadening effects, arising from the nonlinearities of the MZM and the square-law photodiode detection, we apply oversampling inside the transceiver neural networks by a factor of 4 over the 84\,GSa/s DAC rate. The symbol rates of the SBRNN systems become 7\,GSym/s ($n\!=\!48$) and 14\,GSym/s ($n\!=\!24$), for which the corresponding bit rates are 42\,Gb/s and 84\,Gb/s, respectively. The bandwidth of the signal is restricted at both transmitter and receiver by a brick-wall LPF with a cut-off frequency of 32\,GHz. {Note that in practice, down-sampling by a factor of 4 of the filtered series of symbols can be performed without loss of information. Due to the low-pass filtering, the original series of symbols, each of $n\!=\!24$ or $n\!=\!48$ samples at 336\,GSa/s, can be exactly reconstructed from the down-sampled symbol series running at the DAC rate of 84\,GSa/s. Figure~\ref{fig:Schematics} shows the signal spectrum at the transmitter both immediately after the neural network and before the DAC (after the LPF). Because of the LPF, the spectrum of the transmitted signal is confined within 32\,GHz. Down-sampling was performed during the experimental verification of the end-to-end deep learning concept for optical fiber systems reported in~\cite[Sec. V]{Karanov}, but for simplicity, we omitted it in our simulation.} As discussed in Sec.~\ref{sec:Communication_channel}, we set the variance of the additive receiver white Gaussian noise to $\sigma_r^2=2.455\cdot 10^{-4}$. The fiber attenuation and CD parameters are $\alpha\!=\!0.2\,\textnormal{dB/km}$ and $\beta\!=\!17\,\textnormal{ps/nm/km}$, respectively, while the DAC/ADC ENOB is 6. We compare the end-to-end \emph{vanilla} and LSTM-GRU SBRNN designs to the FFNN~\cite{Karanov} as well as PAM2 schemes with multi-symbol FFNN receiver nonlinear equalization (Tx-PAM2 \& Rx-FFNN)~\cite{Bell_labs_OFC2019_Volterra}. In the following paragraph, we briefly describe the latter system, while we refer the interested reader to~\cite{Karanov,Karanov_2} for a thorough description of the end-to-end FFNN system, whose design is identical in this work.
\begin{figure}[t!]
\centering

\includegraphics[]{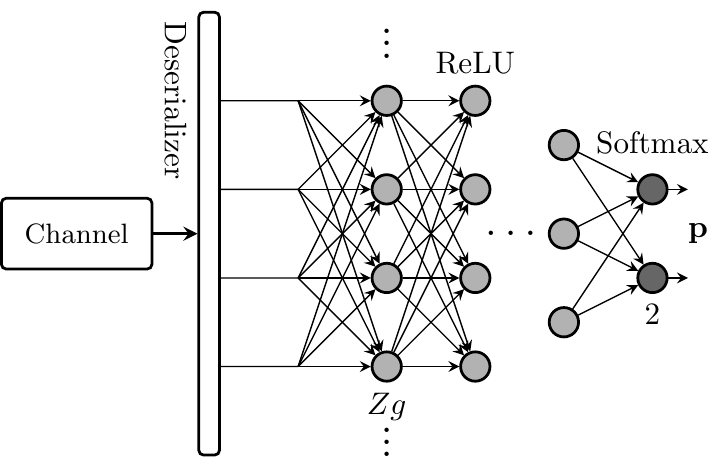}

\caption{\label{fig:Volterra_Schematics} PAM2 system with multi-symbol FFNN receiver as in~\cite{Bell_labs_OFC2019_Volterra}.}
\end{figure}

The PAM2 transmitter directly maps a sequence of bits into a sequence of PAM2 symbols using the levels $\{0;\pi/4\}$. The symbols are pulse-shaped by a raised-cosine (RC) filter with 0.25 roll-off factor. For the 42\,Gb/s system the shaping is performed at $g=8$ samples per symbol to ensure that 48 samples carry 6 bits of information as in the reference SBRNN and FFNN setups. Similarly, for the 84\,Gb/s Tx-PAM2 \& Rx-FFNN system the pulse-shaping is performed at $g=4$ samples per symbol, ensuring 6 bits are carried over 24 samples. The obtained waveform is transmitted over the communication channel described in ~Sec.\ref{sec:Communication_channel}, the first element of which is the 32\,GHz-bandwidth LPF. Figure~\ref{fig:Volterra_Schematics} shows a schematic of the receiver section in the Tx-PAM2 \& Rx-FFNN system. The distorted sequence of samples is received and samples corresponding to multiple neighboring symbols are fed to the receiver multi-layer FFNN for equalization. Thus, the first layer of the deep neural network has parameters $\mathbf{W}_{}\in{\mathbb{R}}^{Zg\times Zg}$ and $\mathbf{b}_{}\in{\mathbb{R}}^{Zg}$, where $Z$ is the symbol processing window of the system. As recommended in~\cite{Bell_labs_OFC2019_Volterra}, multiple hidden layers are employed to further process the received samples before a single symbol is estimated. The number of nodes on each of the hidden layers is given by $\left\lfloor Zg/(2^{l-1})\right\rfloor$, where $l=1$ corresponds to the first hidden layer, $l=2$ is for the second hidden layer, etc.. ReLU activation functions are applied on all layers, except for the final layer which uses the \emph{softmax} activation and outputs a probability vector of length 2 for the corresponding transmitted PAM2 symbol $\mathbf{p}\in{\mathbb{R}}^{2}$. We estimate the central of $Z$ input symbols, thus including preceding and succeeding symbols for receiver processing. After estimation, we slide the processing window one position ahead to estimate the next PAM2 symbol in the sequence. It has been shown that such systems can approximate nonlinear Volterra equalizers~\cite{Bell_labs_OFC2019_Volterra,Lyubomirsky}. Training of the deep FFNN is performed by labeling the transmitted PAM2 sequences and using the cross entropy loss. As with the other systems independent data sequences are generated using the Tausworthe generator in the testing phase. Because of the large size of the receiver FFNN when $Z=61$, we use 400 000 optimization iterations to ensure that training converges.

\begin{table}
 \caption{\textbf{Parameters of the Utilized Neural Networks}}
 \label{table2:Network parameters}
 \centering
 \begin{tabular}{cccc}
 \toprule
 Param. & Tx-PAM2 \& Rx FFNN & FFNN & vanilla/\color{gray}GRU \color{black}SBRNN\\
 \midrule
 Bits per symbol & 1 & $\log_2(M)=6$ & $\log_2(M)=6$ \\
 Samples per symbol & $g=8/4$ & $n=48/24$ & $n=48/24$ \\
 Processing window & $Z=11 / 21 / 61$ & $W=1$ & $W=10$ \\
 Layers & $L = 6 / 7 / 9$ & $3 + 3$ & $2 \cdot (1 + 1) +1$ \\
 \# Nodes & $Zg + \displaystyle\sum_{l=1}^{L-2}\left\lfloor\frac{Zg}{2^{l-1}}\right\rfloor + 2$ & $10M + 2n$ & $\color{black} 15M + 6n/\color{gray}35M + 18n$\\
 \# Nodes for 42\,Gb/s  & $1457$ $(Z=61)$ & $736$ & $ 1248/\color{gray}3104$  \\
 
 \# Nodes for 84\,Gb/s & $728$ $(Z=61)$ & $688$ & $ 1104/\color{gray}2672$  \\

 \bottomrule

 \end{tabular}

\end{table}

Table~\ref{table2:Network parameters} lists the neural network parameters for the SBRNN as well as the Tx-PAM2 \& Rx-FFNN and end-to-end FFNN systems. For the \emph{vanilla} and LSTM-GRU SBRNN, the processing window size is fixed to $W\!=\!10$ blocks during optimization and testing, while for the Tx-PAM2 \& Rx-FFNN we increase the processing window from $Z=11$ to $Z=21$ and then to $Z=61$. Moreover, the number of layers in the Tx-PAM2 \& Rx-FFNN is increased accordingly. In the case of the SBRNN, we use single layer networks at transmitter and receiver. It is important to note that the number of nodes in the proposed SBRNN system does not depend on $W$, while in stark contrast, increasing $Z$ in Tx-PAM2 \& Rx-FFNN results in a tangible increase in the number of nodes. As a consequence, the vanilla SBRNN design at 42\,Gb/s has 209 fewer trainable parameters, while processing the same amount of received samples. The difference becomes even more pronounced if the processing window is increased to capture more of the interference. It should also be mentioned that using \emph{embeddings} instead of \emph{one-hot vector encoding} could potentially further decrease the number of nodes in the end-to-end systems. {As an alternative to performing multiplication with one-hot vectors, embeddings are straightforwardly applied to the FFNN system and we use them in our simulations. For application of embeddings in recurrent neural networks we refer the interested reader to~\cite{RANs}.}
\begin{figure}[t!]
\centering
\includegraphics[]{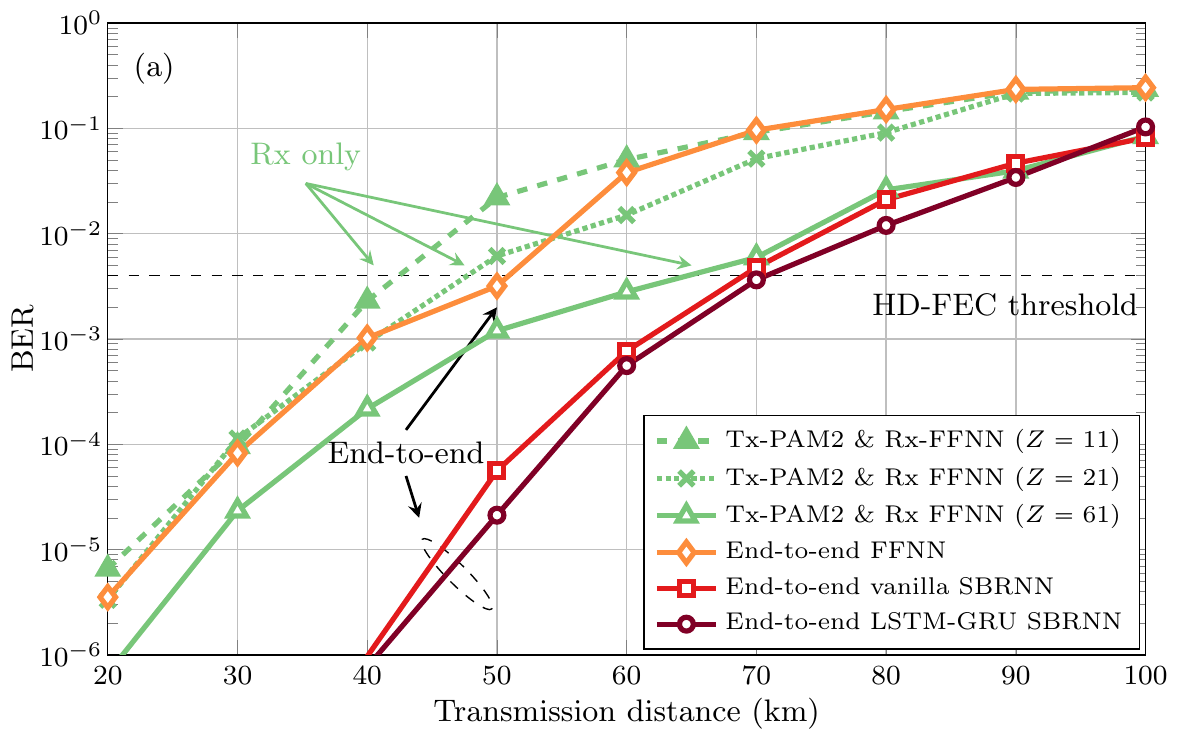}
\includegraphics[width=0.7\textwidth, keepaspectratio=true]{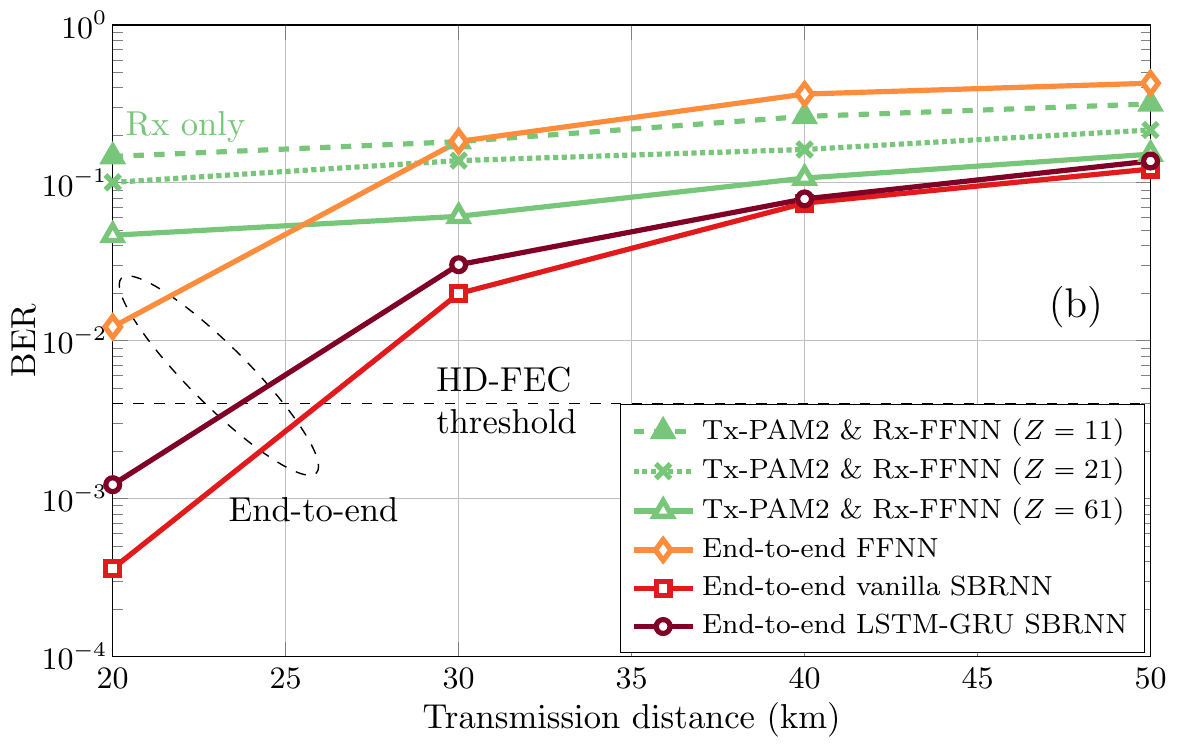}
\includegraphics[width=0.285\textwidth, keepaspectratio=true]{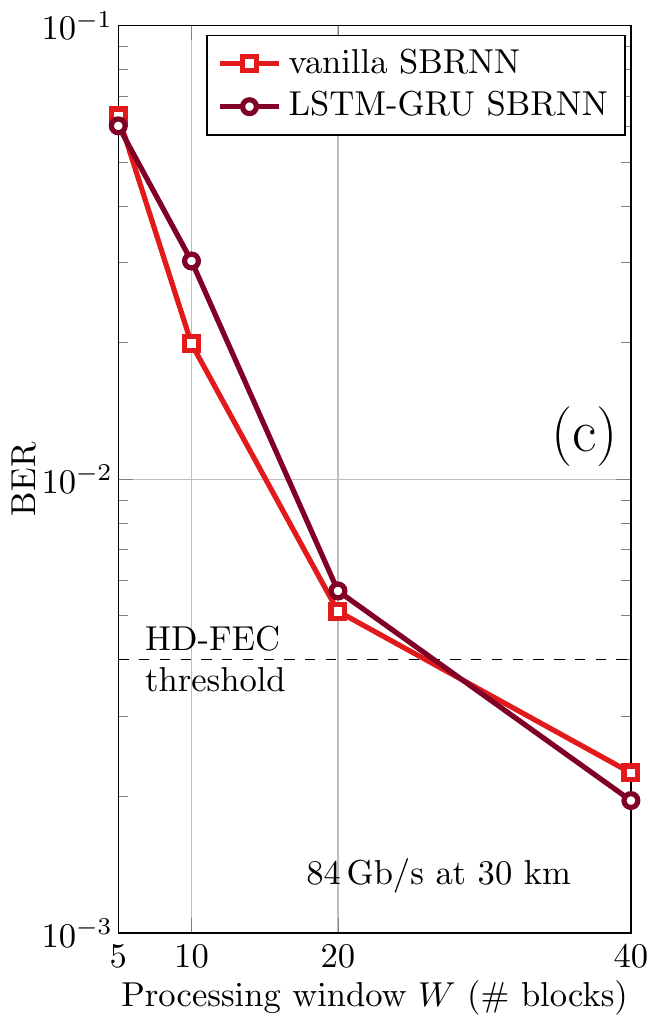}
\caption{BER as a function of transmission distance for the end-to-end \emph{vanilla} and LSTM-GRU SBRNN systems compared to the end-to-end FFNN as well as the PAM2 system with multi-symbol FFNN receiver. The systems operate at (a) 42\,Gb/s and (b) 84\,Gb/s. (c) BER versus processing window for the 84\,Gb/s \emph{vanilla} and LSTM-GRU SBRNN at 30\,km.}
\label{fig:Results}
\end{figure}

Figure~\ref{fig:Results}(a) shows the BER performance of the examined 42\,Gb/s systems at different transmission distances. Note that all systems are trained separately for each distance from 20 to 100\,km (in steps of $10$\,km) and all results represent the average BER achieved by the best parameter sets found in three independent training runs (with different initializations of both the NN parameters and the random number generators). {It should be mentioned that instead of training the neural networks at a fixed nominal distance, alternatively, a novel multi-distance training method can be employed~\cite{Karanov}. It enables the generalization of the neural networks' parameters for operation over a range of link distances, yielding robust and flexible transceivers which do not require reconfiguration. For further details on the multi-distance parameter training for end-to-end FFNN systems, we refer the interested reader to~\cite{Karanov}, while the experimental verification of the method was reported in~\cite{Karanov_2}. The choice of the particular training procedure is a trade-off between system versatility and BER. The application of multi-distance training to systems based on sequence processing, such as the end-to-end SBRNN, falls outside of the scope of this manuscript and is part of an ongoing investigation.} In contrast to the FFNN design, the SBRNN handles data sequences, having a structure that can mitigate inter-block interference during transmission. This allows the SBRNN autoencoders to significantly outperform FFNN at any examined distance. Moreover, our results indicate that the SBRNN designs can enable communication below the $4\cdot10^{-3}$ HD-FEC threshold~\cite{Agrell} at distances of 70\,km, which is a substantial increase over the achievable distance of 50\,km below HD-FEC for the FFNN. The Tx-PAM2 \& Rx-FFNN $(Z=61)$ system, whose receiver simultaneously processes relatively the same number of samples as the SBRNN (488 to 480), is as well outperformed by both the \emph{vanilla} and the LSTM-GRU designs at all distances up to 80\,km. In particular, for distances up to 50\,km both SBRNN systems obtain BERs significantly below the value for the Tx-PAM2 \& Rx-FFNN. Moreover, our simulation prediction indicates that the Tx-PAM2 \& Rx-FFNN $(Z=61)$ cannot achieve transmission below HD-FEC beyond 60\,km, while the \emph{vanilla} and LSTM-GRU SBRNN increase the transmission distance to around 70\,km. At longer distances all three systems achieve BERs above $10^{-2}$. With $Z=61$, the Tx-PAM2 \& Rx-FFNN system performs comparably, however at higher computational cost. Note that with an optimized bit-mapping, the SBRNN systems can be further improved.

Figure~\ref{fig:Results}(b) shows the systems' BER performance at distances from 20 to 50\,km when the data transmission is at 84\,Gb/s. We see that at such rates only the SBRNN systems can achieve reliable communication below HD-FEC at 20\,km, with the BER of the \emph{vanilla} system slightly lower than that of the LSTM-GRU. The obtained BERs for the end-to-end FFNN and PAM2 systems are above $10^{-2}$. Nevertheless, we observe that at 30\,km and beyond, the SBRNN systems cannot obtain BERs below the HD-FEC. This is a consequence of the reduced block size~($n\!=\!24$) which means that the inter-block interference during transmission spans over a greater number of blocks. We extend our investigation to highlight the impact of an increased SBRNN processing window to address the CD effects in the 84\,Gb/s systems. Figure~\ref{fig:Results}(c) shows the BER performance of the 84\,Gb/s \emph{vanilla} and LSTM-GRU at 30\,km for varying window size between 5 and 40 blocks. We see that in both systems, increasing $W$ reduces the BER. In particular an increase to $W=40$ decreases the \emph{vanilla} and LSTM-GRU SBRNN BER to $2.25\cdot 10^{-3}$ and $1.9\cdot 10^{-3}$, respectively, below the HD-FEC threshold. We also note that the gains of increasing the processing window start to diminish for $W=40$, indicating that effects other than ISI, e.g., noise and nonlinearities, become dominant. 

\begin{figure}[t!]
\centering
\includegraphics[]{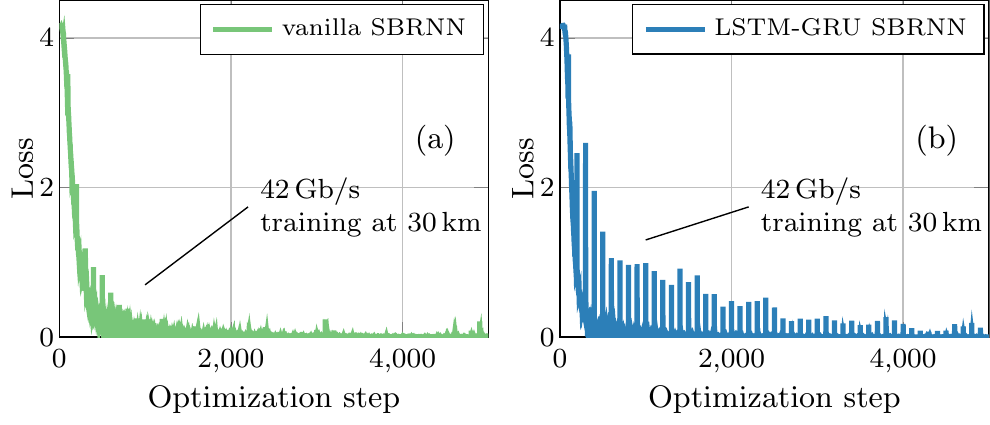}
\caption{Cross entropy loss versus training step for the 42\,Gb/s (a) vanilla and (b) LSTM-GRU SBRNN systems at 30\,km.}
\label{fig:Results_Loss}
\end{figure}

As an interesting observation, it should be mentioned that, although outperformed by the Tx-PAM2 \& Rx-FFNN $(Z=61)$, the end-to-end FFNN system is a design with substantially smaller number of trainable parameters. Moreover, while processing only 48 samples simultaneously at 42\,Gb/s, the system achieves BERs comparable to the Tx-PAM2 \& Rx-FFNN $(Z=11)$ and Tx-PAM2 \& Rx-FFNN $(Z=21)$ systems, which process 88 and 168 samples, respectively. Such a comparison further highlights the advantage of the end-to-end deep learning-based systems, which in contrast to systems with neural network processing only at the receiver, leverage the joint optimization of the message-to-waveform mapping and equalization functions.

It is also worth noting that our results show that the LSTM-GRU SBRNN slightly outperforms the \emph{vanilla} architecture, achieving lower BERs in most scenarios. This suggests that the particular choice of SBRNN design is a trade-off between performance and complexity. We make a direct comparison between the proposed \emph{vanilla} and LSTM-GRU designs, examining the training convergence of two representative 42\,Gb/s systems at 30\,km. The cross entropy loss as a function of optimization step for the \emph{vanilla} and LSTM-GRU SBRNN systems is plotted in  Fig.~\ref{fig:Results_Loss}(a) and \ref{fig:Results_Loss}(b), respectively. We observe peaks in the loss value that occur every 100 optimization steps when the outputs in the forward and backward passes of the BRNN are re-initialized to $\mathbf{0}$ (see Sec.~\ref{transceiver_design}). Moreover, for a fixed value of the loss, the LSTM-GRU structure requires more training iterations. This is an outcome of the increased number of parameters in the LSTM-GRU recurrent structure, which entails a slower training process compared to the \emph{vanilla} SBRNN. In particular, Table~\ref{table2:Network parameters} shows that there are significantly fewer trainable parameters in the \emph{vanilla} SBRNN system  compared to the LSTM-GRU at both 42 and 84\,Gb/s. Yet, the performance of such a design is comparable to that of the LSTM-GRU. This may be accounted to the fact that, as a result of fiber dispersion neighboring symbols impose stronger interference effects. As a consequence, structures such as the LSTM-GRU, designed to capture long term dependencies in a sequence via the sigmoid memory gates, do not significantly improve the performance over \emph{vanilla} RNNs, which consist of a simple concatenation between the current input and the previous output and are sufficient in such applications.

\section{Conclusions}
For the first time,  we propose and investigate end-to-end deep learning of communication over dispersive channels with nonlinear intensity receivers based on bidirectional recurrent neural networks. Such a communication scheme is encountered frequently in short-reach fiber-optic systems. In conjunction with an efficient sequence estimation technique, the proposed SBRNN autoencoder is specifically designed to handle the memory in nonlinear communication channels. Our study shows that we can significantly improve the performance of optical IM/DD communications compared to previous end-to-end deep learning-based systems implemented as feed-forward neural networks (FFNN). In a comparison with PAM2 systems employing state-of-the-art nonlinear equalizers, which operate on multiple received symbols simultaneously, the end-to-end SBRNN allows to increase reach or enhance the data rate for shorter distances, at lower computational complexity. In particular, the LSTM-GRU variant achieves information rates of 42\,Gb/s below the  6.7\% HD-FEC at distances of 70\,km, while at 30\,km, both \emph{vanilla} and LSTM-GRU systems can operate at 84\,Gb/s below the threshold with appropriately chosen parameters. Our work marks an important milestone in the quest of fully unlocking the potential of end-to-end optimized transmission via deep learning by innovating a transceiver design tailored for communication over practical dispersive nonlinear channels.

\section*{Funding}
EU Marie-Sk{\l}odowska-Curie COIN  project (No.676448); CELTIC EUREKA project SENDATE-TANDEM (Project ID C2015/3-2), funded by the German BMBF (Project ID 16KIS0450K).


\begin{thebibliography}{99}

\bibitem{Goodfellow}
I.~Goodfellow, Y.~Bengio and A. Courville, \textit{Deep Learning} (Massachusetts Institute of Technology, 2016).
\bibitem{O'Shea}
T.~O'Shea and J.~Hoydis, ``An introduction to deep learning for the physical layer,'' IEEE Trans. Cogn. Commun. Netw. \textbf{3}(4), 563-575 (2017).
\bibitem{Doerner}
S.~D{\"o}rner, S.~Cammerer, J.~Hoydis, and S.~ten~Brink, ``Deep learning-based communication over the air,'' IEEE J. Sel. Topics Signal Process. \textbf{12}(1), 132-143 (2018).
\bibitem{Karanov}
B.~Karanov, M.~Chagnon, F.~Thouin, T.~A.~Eriksson, H.~B{\"u}low, D.~Lavery, P.~Bayvel, and L.~Schmalen, ``End-to-end deep learning of optical fiber communications,'' J. Lightw. Technol. \textbf{36}(20), 4843-4855 (2018).
\bibitem{Karanov_2}
M.~Chagnon, B.~Karanov, and L.~Schmalen, ``Experimental demonstration of a dispersion tolerant end-to-end deep learning-based IM-DD transmission system,'' in \textit{Proceedings of 44th European Conference on Optical Communications (ECOC)} (Institute of Electrical and Electronics Engineers, 2018), pp. 1-3.
\bibitem{Li}
S.~Li, C.~H{\"a}ger, N.~Garcia, and H.~Wymeersch, ``Achievable information rates for nonlinear fiber communication via end-to-end autoencoder learning,'' in \textit{Proceedings of 44th European Conference on Optical Communications (ECOC)} (Institute of Electrical and Electronics Engineers, 2018), pp. 1-3.
\bibitem{Jones}
R.~Jones, T.~A.~Eriksson, M.~P.~Yankov, B.~J. Puttnam, G.~Rademacher, R.~S.~Luis, and D.~Zibar, ``Geometric constellation shaping for fiber optic communication systems via end-to-end learning,'' \textit{ArXiv preprint arXiv:1810.00774} (2018).
\bibitem{Agrawal}
G.~Agrawal, \textit{Fiber-optic Communication Systems, 4th ed.} (John Wiley \& Sons, Inc., 2010).
\bibitem{Rumelhart}
D.~Rumelhart, G.~Hinton, and R.~Williams, ``Learning representations by back-propagating errors,'' Nature \textbf{323}, 533-536 (1986).
\bibitem{Ye}
C.~Ye \textit{et al.}, ``Recurrent neural network (RNN) based end-to-end nonlinear management for symmetrical 50Gbps NRZ PON with 29dB+ loss budget,''  in \textit{Proceedings of 44th European Conference on Optical Communications (ECOC)} (Institute of Electrical and Electronics Engineers, 2018), pp. 1-3.
\bibitem{Schuster}
M.~Schuster and K. Paliwal, ``Bidirectional recurrent neural networks,'' IEEE Trans. Signal Process. \textbf{45}(11), 2673-2681 (1997).
\bibitem{Farsad}
N.~Farsad and A.~Goldsmith, ``Neural network detection of data sequences in communication systems,'' IEEE Trans. Signal Process. \textbf{66}(21), 5663-5678 (2018).
\bibitem{Farsad_2}
N.~Farsad and A.~Goldsmith, ``Neural network detectors for molecular communication systems,'' in \textit{Proceedings of 19th International Workshop on Signal Processing Advances in Wireless Communications (SPAWC)} (Institute of Electrical and Electronics Engineers, 2018), pp. 1-5.
\bibitem{Cho}
K.~Cho, B.~van~Marrienboer, C.~Gulcehre, D.~Bhadanau, F.~Bougares, H.~Schwenk, and Y.~Bengio, ``Learning phrase representations using RNN encoder-decoder for statistical machine translation,'' in \textit{Proceedings of Conference on Empirical Methods in Natural Language Processing} (Association for Computational Linguistics, 2014), pp. 1724-1734.
\bibitem{Ravanelli}
M.~Ravanelli, P.~Brakel, M.~Omologo, and J.~Bengio, ``Light gated recurrent units for speech recognition,'' IEEE Transactions on Emerging Topics in Computational Intelligence \textbf{2}(2), 92-102 (2018).
\bibitem{Hochreiter}
S.~Hochreiter, and J.~Schmidhuber, ``Long short-term memory,'' Neural Computation \textbf{9}(8), 1735-1780 (1997).
\bibitem{Bell_labs_OFC2019_Volterra}
V. Houtsma, E. Chou, and D. van Veen, ``92 and 50 Gbps TDM-PON Using Neural Network Enabled Receiver Equalization Specialized for PON,'' in \textit{Proceedings of Optical Fiber Communication Conference (OFC)} (Optical Society of America, 2019), paper M2B.6.
\bibitem{Lyubomirsky}
I.~Lyubomirsky, ``Machine learning equalization techniques for high speed PAM4 fiber optic communication systems,'' CS229 Final Project Report, Stanford University (2015). Accessed on: Mar. 13, 2019. [Online]. Available: http://cs229.stanford.edu/proj2015/232\_report.pdf.
\bibitem{Nair}
V.~Nair and G.~Hinton, ``Rectified linear units improve restricted Boltzmann machines,'' in \textit{Proceedings of International Conference on Machine Learning (ICML)} (International Machine Learning Society, 2010), pp.~807-814.
\bibitem{Kingma}
D.~Kingma and J.~Ba, ``Adam: A method for stochastic optimization,'' \textit{ArXiv preprint arXiv:1412.6980} (2014).
\bibitem{Tensorflow}
Tensorflow, https://www.tensorflow.org/.
\bibitem{Eriksson}
T.~A.~Eriksson, H.~B{\"u}low, and A.~Leven, ``Applying neural networks in optical communication systems: possible pitfalls,'' IEEE Photon. Technol. Lett. \textbf{29}(23), 2091-2094 (2017).
\bibitem{Lee}
D.-U~Lee, J.~Villasenor, W.~Luk, and P.~Leong, ``A hardware Gaussian noise generator using the Box-Muller method and its error analysis,'' IEEE Trans. Comput. \textbf{55}(6), 659 (2006).
\bibitem{RANs}
K.~Lee, O.~Levy, and L.~Zettlemoyer, ``Recurrent additive networks,'' \textit{ArXiv preprint arXiv:1705.07393} (2017).
\bibitem{Agrell}
E.~Agrell and M.~Secondini, ``Information-theoretic tools for optical communication engineers,'' in \textit{Proceedings of IEEE Photonics Conference (IPC)} (Institute of Electrical and Electronics Engineers, 2018), pp. 1-5.
\end{thebibliography}
\end{document}